\documentstyle[12pt]{article}
\begin{document}
\textwidth=17cm
\textheight=22cm
\topskip=-1cm
\footskip=1cm
\footheight=1cm
\oddsidemargin=-1cm %marge de gauche

%\title{ Cosmic Microwave Background Dipole induced by double inflation}
%\author{ David Langlois}
%\address{}
\begin{center}
{\bf Cosmic Microwave Background Dipole induced by double inflation }\\
\vskip .5cm
{\bf David Langlois } \\
{\it D\'epartement d'Astrophysique Relativiste et de Cosmologie,\\
Centre National de la Recherche Scientifique,\\ Observatoire de
Paris, 92195 Meudon, France\\
and\\
 Racah Institute of Physics, The Hebrew University,\\
 Givat Ram, 91904 Jerusalem, Israel.}\\
\end{center} 

%\maketitle
 
\par\bigskip

\begin{abstract}
The observed CMBR dipole is generally interpreted as the consequence 
of the peculiar motion of the Sun with respect to the reference frame
of the CMBR. This article  proposes an alternative interpretation in which the 
observed dipole is the result of isocurvature perturbations on scales 
larger than the present Hubble radius. These perturbations are produced 
in the  simplest model of double inflation, depending on  three 
parameters. The observed dipole and quadrupole can be explained in this 
model, while  severely  constraining its  parameters.
\end{abstract}

\par\bigskip

The dipole moment is the most distinctive  feature in
the  Cosmic Microwave Background Radiation (CMBR) anisotropy \cite{Dipole}.
It is larger than the quadrupole, measured recently by the satellite
COBE,  
by two orders of magnitude. In general, this dipole is interpreted 
as the peculiar motion of the Sun with respect to the CMBR ``rest 
frame'', whereas the other multipoles are seen as a consequence of 
primordial cosmological perturbations of a homogeneous and isotropic 
universe. 

The satellite COBE has measured a dipole corresponding to a velocity
of $369 \pm 3$ km s$^{-1}$ (\cite{kogutetal}).
 Substracting  the motion of the sun in the galaxy and  the motion of
our galaxy with respect to the local group, one ends up  with a net 
dipole, which, if interpreted as a Doppler effect, yields
   a velocity for the local group of $627
\pm 22$ km s$^{-1}$  towards the galactic coordinates
($l=276^o \pm 3^o, b=33^o \pm 3^o$).
However, although rarely mentioned, the measured dipole is {\it not 
necessarily} 
due to a  Doppler effect  (\cite{pp},\cite{turner}, \cite{lp}). 
If the dipole is only due to  our peculiar motion, then one should 
expect a similar dipole in other cosmological observations.
But such observations are inconclusive up to now: whereas the 
measured dipoles  from  the X-ray background \cite{boldt}, from nearby
galaxies \cite{llb}, from IRAS galaxies \cite{Strauss} and from  distant
supernovae \cite{rpk} are compatible with  the COBE
observation  within the statistical errors,   the observed dipole in 
distant Abell clusters \cite{lap} ($561 \pm 284$ km s$^{-1}$ towards
($l=220^o \pm 27^o, b=-28^o \pm 27^o$))  is
inconsistent with the CMBR dipole.
One cannot therefore reject the possibility that the CMBR dipole is 
incompatible with our peculiar motion. 

 Recently, Langlois and Piran \cite{lp}  have shown that the  CMBR dipole 
could be the result of  isocurvature perturbations on 
scales  larger than the current Hubble radius, but
{\it only with wavelengths bigger than a critical 
scale of the order of a hundred times the Hubble radius today} to avoid  
a quadrupole incompatible with observations. Observationally, the 
measured CMBR dipole would then be the vectorial sum of the isocurvature 
dipole plus the peculiar motion dipole. It could thus be completely different
from the other cosmological  dipoles, which are sensitive only to our 
peculiar motion.

The purpose of this article is to propose a scenario 
that explains the {\it origin}  of isocurvature  perturbations with such an 
abrupt lower cut-off in their spectrum. It is  based on the simplest model
 of multiple inflation: two non-interacting massive scalar fields.
Although most of the pioneering works on inflation considered only one 
scalar field, it was noticed early that inflation with several scalar fields
(motivated by various models of grand unification in particle physics)  
could induce {\it isocurvature} perturbations (see e.g. \cite{kl}).
However, the original motivation for ``double inflation'' (\cite {st})
was to taylor a non-flat  spectrum of {\it adiabatic} 
perturbations that could reconcile various observations of large-scale 
structures. This spectrum  was recently reinvestigated in great detail
by Polarski and Starobinsky (\cite{ps1}). 

Here, the very same ingredients of double inflation are going to be used, 
but with a special emphasis on isocurvature perturbations, and most 
essentially, by assuming that the break in double inflation, i.e.
the transition between inflation driven by the heavy scalar field 
and inflation driven by the light scalar field, takes place at  scales
corresponding today to scales far larger than the Hubble radius. 

The simplest model of double inflation, consisting of two
minimally coupled massive scalar fields, is  described by the Lagragian
\begin{equation} 
{\cal L}={{}^{(4)}
R\over 16\pi G}-{1\over 2} \partial_\mu\phi_l\partial^\mu\phi_l
 -{1\over 2} m_l^2\phi_l^2 -{1\over 2} \partial_\mu\phi_h\partial^\mu\phi_h
 -{1\over 2} m_h^2\phi_h^2,
\end{equation}
where $\phi_l$ and $\phi_h$ are respectively the light and heavy scalar
fields ($m_l < m_h$); ${}^{(4)}R$ is the scalar spacetime curvature and $G$ is 
Newton's constant.
The equations of motion in a flat Friedmann-Lemaitre-Robertson-Walker (FLRW)
spacetime, with the metric $ds^2=-dt^2+a^2(t)d{\bf x}^2$,
 are given by
\begin{equation}
3H^2=4\pi G\left(\dot \phi_l^2+\dot \phi_m^2
+m_l^2\phi_l^2+m_h^2\phi_h^2\right),\label{energyconstraint}
\end{equation}
\begin{equation}
\ddot\phi_l+3H\dot\phi_l+m_l^2\phi_l=0, \qquad
\ddot\phi_h+3H\dot\phi_h+m_h^2\phi_l=0, \label{eqm}
\end{equation}
where a dot denotes a derivative with respect to the cosmic time $t$, and
$H\equiv \dot a /a$ is the Hubble parameter.  
One assumes initial conditions such that $m_h \ll H$; 
$\phi_h,\phi_l \gg m_P/\sqrt{2\pi}$ 
( $m_P\equiv G^{-1/2}$ is the Planck mass), 
so that  one begins with  inflation ($|\dot H|\ll H^2$)
 where  both  scalar fields are slow-rolling. 

Following the analysis of \cite{ps1}, the two scalar fields,
in the slow-rolling approximation
 ($\dot\phi_l^2$ and $\dot\phi_h^2$ are neglected in (\ref{energyconstraint}),
$\ddot\phi_l$ and $\ddot\phi_h$  in (\ref{eqm})),  can be
written in the form
\begin{equation}
\phi_h^2={s\over 2\pi G}\sin^2\theta, \qquad
\phi_l^2={s\over 2\pi G}\cos^2\theta, \label{param}
\end{equation}
where $s=-ln(a/a_e)$ (``e'' stands for the end of inflation).
 Their  evolution is given parametrically by  \cite{ps1}
\begin{equation}
s=s_0\left[{(\sin\theta)^{m_l^2}\over
      (\cos\theta)^{m_h^2}}\right]^{2\over m_h^2-m_l^2}, \quad
H^2(s)={1\over 3}s\left[m_h^2+m_l^2-(m_h^2-m_l^2)\cos 2\theta(s)\right],
\label{H(s)}
\end{equation}
where $s_0$ is a constant of integration. 

Assume now $m_h/m_l \gg 1$. In this model, several epochs 
can be distinguished. It will be  convenient to label them according  
to the value of the  comoving Hubble length scale, defined by 
$\lambda(t)= (a(t)H(t))^{-1}$.
The initial period of inflation driven   by the heavy scalar field
lasts  until equality of the 
energies of the two scalar fields ($m_h\phi_h=m_l\phi_l$), which  occurs for 
$\theta\simeq m_l/m_h$ and $s\simeq s_0$. The Hubble scale at that time will
be denoted $\lambda_b$ (for ``break'').
If the condition $s_0>(m_h/ m_l)^2$
is satisfied, which will be assumed here,  inflation goes on,
 now driven by the light scalar field (if not,
 there is  a dust-like transition
period before  the second phase of inflation, see \cite{ps1} for details).
Meanwhile the heavy scalar field  continues its slow-rolling until $H\sim m_h$,
after which it oscillates. The Hubble scale corresponding to the end
of the slow-rolling of 
 $\phi_h$ will be denoted  $\lambda_c$ (for ``critical'').
Finally  the Hubble parameter reaches the value $m_l$ 
and inflation stops. 
In the ``standard''  model of double inflation, the scales $\lambda_b$
and $\lambda_c$ (one always has $\lambda_b>\lambda_c$)  are within the 
Hubble radius today, denoted $\lambda_H$, (\cite{st}, \cite{ps1}) 
whereas, here, the parameters 
of the model are chosen such that these two scales are {\it outside} 
the present Hubble radius $\lambda_H$.

Let us now study the perturbations around the background solution given above. 
In the longitudinal gauge the scalar perturbations of the metric reduce to the 
 ``relativistic gravitational potential''  perturbation 
 $\Phi$, so that the metric 
is given by
\begin{equation}
ds^2=-(1+2\Phi)dt^2+a^2(t)(1-2\Phi)\delta_{ij}dx^idx^j.
\end{equation}
The perturbations for the scalar fields are denoted $\delta\phi_h$ and 
$\delta\phi_l$.

From now on, all the perturbations will be decomposed into Fourier modes, 
defined for any field $f({\bf x})$ by 
\begin{equation}
f_{\bf k}=\int {d^3{\bf x}\over (2\pi)^{3/2}}e^{-i {\bf k}.{\bf x}}f({\bf x}),
\end{equation}
and the subscript ${\bf k}$ will be most of the time implicit.

In the limit of small $k$, there exists 
a  solution for $\Phi$ corresponding to the adiabatic growing mode 
 which is independent of the specific matter content, and can be expressed, 
quite generally, as
\begin{equation}
\Phi=C_1\left(1-{H\over a}\int_0^t adt'\right),\label{adiab}
\end{equation}
where $C_1({\bf k})$ is time independent.

The perturbed Einstein equations yield the evolution equations for the 
perturbations, which read:
\begin{equation}
\dot\Phi+H\Phi=4\pi G\left(\dot\phi_h\delta\phi_h
+\dot\phi_l\delta\phi_l\right),
\end{equation}
\begin{equation}
\ddot{\delta\phi_h}+3H\dot{\delta\phi_h}+\left({k^2\over a^2}+m_h^2\right)\delta\phi_h=
4\dot\phi_h\dot\Phi-2m_h^2\phi_h\Phi, \label{eqmpertl}
\end{equation}
with a similar equation for $\delta\phi_l$. 
When the two scalar fields are slow-rolling, the dominant solutions 
for the long wavelength modes ($k\ll aH$) are given by 
\begin{equation}
\Phi=-{C_1\dot H\over H^2}+2C_3{(m_h^2-m_l^2)m_h^2\phi_h^2m_l^2\phi_l^2\over
3(m_h^2\phi_h^2+m_l^2\phi_l^2)^2},
\end{equation}
\begin{equation}
{\delta\phi_l\over\dot\phi_l}={C_1\over H}-2C_3{Hm_h^2\phi_h^2\over
m_h^2\phi_h^2+m_l^2\phi_l^2},\quad
{\delta\phi_h\over\dot\phi_h}={C_1\over H}+2C_3{Hm_l^2\phi_l^2\over
m_h^2\phi_h^2+m_l^2\phi_l^2},\label{perth}
\end{equation}
where $C_1({\bf k})$ (the same as in (\ref{adiab}))
 and $C_3({\bf k})$ are time independent. 
The terms with $C_1$ will give later (in the radiation era)
 the adiabatic growing mode, whereas the terms proportional to $C_3$
will give the non-decaying isocurvature mode.

The initial conditions are obtained from the vacuum quantum fluctuations of 
the two independent fields $\delta\phi_h$ and $\delta\phi_l$. 
 Following the same procedure as for inflation with one scalar field, one 
finds that $\delta\phi_h$ and $\delta\phi_l$
can be represented, for wavelengths crossing out the Hubble radius, 
 as classical random fields such that
\begin{equation}
\delta\phi_h={H(t_k)\over\sqrt{2k^3}}e_h({\bf k}), \qquad
\delta\phi_l={H(t_k)\over\sqrt{2k^3}}e_l({\bf k}), 
\label{quantization}
\end{equation}
where $e_h$ and $e_l$ are centered and  normalized gaussian fields, 
i.e. $\langle e_h({\bf k})\rangle=0$, $\langle e_h({\bf k})e_h({\bf k}')\rangle=
\delta({\bf k}-{\bf k}')$ (and the same for $e_l$); $t_k$ is the time when the
wavelength crosses the Hubble radius, i.e. when $k=2\pi a H$.
Inverting (\ref{perth}) then defines $C_1$ and $C_3$ as random gaussian fields.

During the matter era, $\Phi=(3/5)C_1$, which follows from the general 
solution (\ref{adiab}).
Defining  the spectrum of any homogeneously
and isotropically distributed random field $f({\bf x})$ by 
$\langle f_{\bf k} f_{\bf k}'\rangle=2\pi^2k^{-3}{\cal P}_f(k)\delta(
{\bf k}-{\bf k}')$,
 the spectrum for $\Phi$ today, can be found \cite{ps1} to be, 
using (\ref{H(s)}):
\begin{equation}
{\cal P}_\Phi={6 G\over 25\pi}s^2\left[m_h^2+m_l^2-(m_h^2-m_l^2)\cos 2\theta(s)
\right]. \label{spectrum}
\end{equation}
$s$ is taken at the time $t_k$, and since the variation of $H$ during 
inflation is very small with respect 
to that of the scale factor, $s(k)\equiv s(t_k)\simeq \ln (k_e/k)$.
This spectrum has two plateaus, as explained in \cite{ps1}. 
Since, here, the transition scale is larger than the Hubble radius 
today, only the lower plateau, corresponding to the second phase of 
inflation (driven by the light scalar field), and given by
 the limit $\theta=0$ of (\ref{spectrum}),
\begin{equation}
{\cal P}_\Phi={12\over 25\pi}s^2{m_l^2\over m_P^2},
\end{equation}
 will be accessible to 
observations.  The most direct observational consequence of inflation-generated
perturbations is the CMBR fluctuations at large angular scales. At these
scales, the fluctuations are due mainly to the Sachs-Wolfe effect (\cite{sw}): 
\begin{equation}
\left({\Delta T\over T}\right)_{SW}(\vec e)={1\over 3}\left[\Phi_{ls}-\Phi_0
\right]+\vec e.\left[\vec v_{ls}-\vec v_0\right], \label {sw}
\end{equation} 
 where $\vec e$ gives the direction of
observation on the celestial sphere.  The subscript $ls$ refers to the last
scattering surface and the subscript $0$ to the observer today. 
Decomposing the  CMBR fluctuations  in
spherical harmonics, the prediction for the harmonic components  
$l\ge 2$  on large angular scales, 
due to  the Sachs-Wolfe effect, is
\begin{equation}
\Sigma_l^2=\langle|a_{lm}|^2\rangle={4\pi\over 9} \int {dk\over k} 
{\cal P}_\Phi(k)  j_l^2(2k/a_0H_0).  \label{sigma}
\end{equation}
This formula is obtained by implicitly assuming that the 
contribution of the isocurvature perturbations to the harmonic 
components  $l\ge 2$
is negligible (this is in general required for compatibility between CMBR
and large scale structure observations, see e.g. \cite{ll}). 
The spectrum depends on the scale only logarithmically, and the 
multipoles can thus be obtained, with a good approximation,
 by using the expression for 
a flat spectrum:
\begin{equation}
a_l^2\equiv {2l+1\over 4\pi}\Sigma_l^2=
{6\over 225 \pi}\left({m_l\over m_P}\right)^2{2l+1\over l(l+1)}s^2_l,
\label{quadrupole}
\end{equation}
where $s_l$ is the value for $s$ corresponding to $k=la_0H_0/2$, 
 which contributes the most to the integral (\ref{sigma}).
The quadrupole $a_2=(m_l/ m_P)s_H/(3\sqrt{5\pi})$ 
 must be adjusted to agree with the COBE measurement
 of $Q_{rms-PS}(n=1)\simeq 18 \mu K$ (\cite{banday}).
Hence the constraint on the model:
\begin{equation}
\left({m_l\over m_P}\right) s_H \simeq 7.8\times 10^{-5}.
\end{equation}
If one takes $s_H=60$, this gives $m_l\simeq 1.3 \times 10^{-6} m_P$.

In addition to adiabatic pertubations, this model can also produce 
 isocurvature perturbations. This is the case for example if one assumes
that the light scalar field, after inflation, will decay into ordinary 
matter, while the heavy scalar field remains decoupled from ordinary 
matter and will contribute to the cold dark matter. After inflation, there
will then be two species: ordinary matter 
which behaves as radiation (its background energy density is 
denoted $\rho_r$ and the perturbation $\delta\rho_r$) and 
 the heavy scalar field particles, which behave as dust-like matter (
with  energy density  $\rho_m +\delta\rho_m$). 
One must be careful to use 
comoving energy density perturbations, 
i.e.  with respect to the frame
where matter is at rest (see \cite{lp}). Then the isocurvature perturbation 
is defined
by
\begin{equation}
S=\delta_m -{3\over 4}\delta_r,
\end{equation}
with $\delta_m\equiv \delta\rho_m/\rho_m$,  $\delta_r\equiv \delta\rho_r/\rho_r$.
 For the  scalar field $\phi_h$, the comoving energy density perturbation 
is given by
\begin{equation}
\delta\rho_h=\dot\phi_h\dot{\delta\phi_h}+m_h^2\phi_h\delta\phi_h+3H\dot\phi_h\delta\phi_h
-\dot\phi_h^2\Phi.
\end{equation}
 During inflation dominated by the light scalar field,
the heavy scalar field still slow rolling,  eq (\ref{perth}) implies that
$\delta\phi_h=-{2\over 3}C_3 m_h^2\phi_h \label{dphih}$
(for the modes $k\ll aH$).
Moreover, $\delta\phi_h$ satisfies the same equation as that for the background
$\phi_h$ (because  one can neglect the right-hand side of 
eq (\ref{eqmpertl})) 
and one can thus use the constancy of $\delta\phi_h/\phi_h$ to 
match various periods. In particular, during the oscillating phase, $\phi_h
\sim\delta\phi_h\sim a^{-3/2}\sin(m_h t+\gamma)$, and therefore 
$\delta_m= \delta\rho_h/ \rho_h=2\delta\phi_h/\phi_h=-(4/3)m_h^2C_3$.
During the radiation era, $S\simeq\delta_m$ and 
 the spectrum of $S$, which was given in \cite{ps2}, then follows from
(\ref{perth}) and (\ref{quantization}):
\begin{equation} 
{\cal P}_S={H^2\over \pi^2}\left[{1\over\phi_h^2}+{m_h^4\over m_l^4\phi_l^2}
\right].
\end{equation}
When $\theta \ll 1$, the second term on the right hand side can be neglected
and one finds, using (\ref{param}) and (\ref{H(s)}), 
\begin{equation} 
{\cal P}_S={4 G\over 3\pi}m_l^2\left({s_0\over s}\right)^{m_h^2/m_l^2}.
\end{equation}
This expression is valid for scales crossing out  the Hubble radius while 
$\phi_h$ is still slow-rolling, i.e. for $\lambda>\lambda_c$.
One notices that the spectrum increases when $k$ increases. However, when 
$\phi_h$ stops slow-rolling, the production of isocurvature perturbations
falls abruptly. This occurs when $H\sim m_h$, corresponding to $s=s_c 
\sim m_h^2/m_l^2$, and the isocurvature spectrum then reaches its maximum.

As emphasized in \cite{lp}, the isocurvature modes remain constant during
the evolution of the universe, {\it as long as they remain outside the 
Hubble radius}. This is the case for the modes that contribute to the 
dipole and the quadrupole of the CMBR. Therefore, the previous spectrum 
produced during inflation will remain the same until the last scattering, 
for the large scales. 

Let us now consider the dipole generated by the pertubations larger than the 
 present Hubble radius. As shown in 
\cite{lp} the crucial difference between the dipole and the other multipoles
is that, for the dipole, the terms $\Phi /3$ and ${\bf e}.{\bf v}$ in the Sachs-Wolfe
contribution will cancel each other. Therefore, even if the adiabatic 
perturbations are dominant for the multipoles $l\ge 2$, the isocurvature 
perturbations will dominate  the dipole, because of their contribution 
to the intrinsic fluctuations
\begin{equation}
\left({\Delta T\over T}\right)_{int}
\simeq -{1\over 3}S.
\label{intriso}
\end{equation}
The corresponding expected dipole is thus
\begin{equation}
a_1^2\equiv {3\over 4\pi} \Sigma_1^2={1\over 3} \int {dk\over k} 
{\cal P}_S(k)  j_1^2(2k/a_0H_0),  
\end{equation}
Introducing the variable 
$X=\ln (k_c/ k)$, this  can be expressed more conveniently as
\begin{equation}
a_1^2= {16\over 81\pi}\left({m_l\over m_P}\right)^2 e^{2(s_H-s_c)}
\left({s_0\over s_c}\right)^{m_h^2/m_l^2}
\int_0^\infty dX {e^{-2X}\over (1+X/s_c)^{m_h^2/m_l^2}}, \label{dipole}
\end{equation}
 where has been used the fact that $j_1(x)\sim x/3$ for small $x$.
The integral in (\ref{dipole}) 
is well approximated as $(2+(m_h/m_l)^2/s_c)^{-1}$.
For consistency, one must also check that the quadrupole 
generated by isocurvature perturbations
is smaller than the adiabatic quadrupole given in (\ref{quadrupole}).
This requires, as seen in \cite{lp}, that 
\begin{equation}
s_c >  s_H + 4.6, \label{cond}
\end{equation}
i.e. that the cut-off scale $\lambda_c$ is one hundred times larger than
the Hubble radius today.
The value $s_H$ is of the order of $60$. Its exact value 
depends on the precise history of the universe since the end of inflation,
as well as the value of the Hubble parameter at that time. 
 Because of the power law dependence
in (\ref{dipole}), {\it  the dipole can be  much bigger 
than the quadrupole}. 
 Because of the condition (\ref{cond}), implying that $m_h^2/m_l^2$
must be large, the main constraint, surprisingly,  comes
from requiring that the dipole is not too high with respect to the 
other multipoles.  $s_0/s_c$ must be very close to 1, which implies
 that the two remaining free parameters of the model, $s_0$
and $m_h$,  must be fine-tuned  according to the relation
\begin{equation}
\left({s_0\over s_c}\right)^{m_h^2/2 m_l^2}
\sim \left({a_1\over a_2}\right) s_H e^{s_c -s_H},
\end{equation}
with $s_c\sim m_h^2/m_l^2$. The higher $s_H$, the more stringent the 
fine-tuning must be.

Note finally that the scenario presented here is not incompatible with  
the standard double inflation scenario, where the break between the 
two phases of inflation occurs at a scale well within the present Hubble 
radius: they can coexist  if one considers three scalar fields, the 
intermediate (in mass)  scalar field playing the role of the light scalar
field in the scenario proposed here and of  the massive scalar field in
standard double inflation.

In conclusion,  whereas previous alternative interpretations for the 
dipole invoke {\it preexisting} cosmological perturbations, 
 the model proposed here  is the first, to my knowledge, to
 provide a mechanism that {\it creates} these perturbations.  
It is possible that the  required   fine-tuning  could
 be relaxed in a model more 
sophisticated  than one with two non-interacting massive scalar fields. 
In any case, this work  shows  that the standard  Doppler 
interpretation should not be taken 
for  granted, and that further observational investigation is needed 
to decide conclusively about the origin of the CMBR dipole.

\end{document}